\documentclass[12pt]{iopart}
\usepackage{amssymb}
\usepackage{epsfig}
\usepackage{subfigure}
\usepackage{graphicx}
\usepackage{textcomp}
\usepackage{cite}
\usepackage{float}

\expandafter\let\csname equation*\endcsname\relax
\expandafter\let\csname endequation*\endcsname\relax
\usepackage{amsmath}
\newcommand{\agt}{\mathrel{\raise.3ex\hbox{$>$\kern-.75em\lower1ex\hbox{$\sim$}}}}
\begin{document}

\title{Emergence of Clustering in an Acquaintance Model without Homophily}

\author{Uttam Bhat} 
\address{Department of Physics, Boston University, Boston, MA 02215, USA}
\address{Santa Fe Institute, 1399 Hyde Park Road, Santa Fe, New Mexico 87501, USA}
\author{P. L. Krapivsky}
\address{Department of Physics, Boston University, Boston, MA 02215, USA}
\author{S. Redner}
\address{Santa Fe Institute, 1399 Hyde Park Road, Santa Fe, New Mexico 87501, USA}
\address{Center for Polymer Studies and Department of Physics, Boston University, Boston, MA 02215, USA}

\begin{abstract}
  We introduce an agent-based acquaintance model in which social links are
  created by processes in which there is no explicit homophily.  In spite of
  the homogeneous nature of the social interactions, highly-clustered social
  networks can arise.  The crucial feature of our model is that of variable
  transitive interactions.  Namely, when an agent introduces two unconnected
  friends, the rate at which a connection actually occurs between them
  depends on the number of their mutual acquaintances.  As this transitive
  interaction rate is varied, the social network undergoes a dramatic
  clustering transition.  Close to the transition, the network consists of a
  collection of well-defined communities.  As a function of time, the network
  can also undergo an \emph{incomplete} gelation transition, in which the
  gel, or giant cluster, does not constitute the entire network, even at
  infinite time.  Some of the clustering properties of our model also arise,
  but in a more gradual manner, in Facebook networks.  Finally, we discuss a
  more realistic variant of our original model in which there is a soft
  cutoff in the rate of transitive interactions.  With this variant, one can
  construct network realizations that quantitatively match Facebook networks.

\end{abstract}
\maketitle

\section{Introduction to Acquaintance Modeling}
\label{intro}

An important feature of many complex networks is that they can be highly
clustered.  That is, such networks are comprised of well-connected modules,
or communities, with weaker connections between them (see, e.g.,
\cite{GN02,NG04,CNM04,RCCLP04,PDF05,N06,BGLF08,CJK09,F10}).  Part of the
motivation for focusing on communities is that unraveling this substructure
may provide important clues about how such networks are organized, how they
function, and how information is transmitted across them.  While identifying
communities has become a standard diagnostic of
networks~\cite{ADGG04,CSGC05,LLDM09,LF11}, and there has been much recent
effort devoted to determine the community structure of complex networks, less
is known about mechanisms that could lead to this heterogeneity; for
contributions in this direction, see, e.g.,
\cite{BPDA04,LM05,TOSHK06,LA07a,LA07b}.  Our goal is to develop a basic model
for the formation of a social network in which highly-clustered substructures
emerge spontaneously from homogeneous social interaction rules.  Thus there
is no need to appeal to homophily (see, e.g., \cite{CGEM06,GJ09}) or some
other explicit source of heterogeneity to generate large-scale clustering.

In our modeling, the starting network consists of $N$ complete strangers with
no links between them.  This might describe, for example, a set of entering
students to a university in an unfamiliar location.  We assume that the
population remains constant over the time scale that social connections form.
There are two distinct ways that connections are made:
\begin{itemize}
\item \emph{Direct Linking:} An agent with either zero friends or one friend
  links to a randomly selected agent.
\item \emph{Transitive Linking:} An agent with two or more friends introduces
  two of them at random.  These selected agents then create a link with a
  rate that is specified below.
\end{itemize}
These two mechanisms underlie the acquaintance model that was introduced by
Davidsen et al.~\cite{DEB02}; for related work see \cite{MVRS04,KT13} and
references therein.  In \cite{DEB02}, the rates of these two linking
processes were fixed and a steady state was achieved by allowing any agent
and all its attached links to disappear at a (small) rate and correspondingly
adding a new agent to the network to keep the number of agents fixed.  In
this work, we impose a different, socially-motivated, mechanism that allows
the network to reach a non-trivial long-time state.

A natural reason for distinguishing between direct and transitive linking is
that an individual with many friends typically has less impetus to initiate
additional connections.  Indeed, it has been previously noted that there
seems to be a cognitively-limited upper limit---the Dunbar number, which is
of the order of a hundred---for the number of meaningful friends that any
individual can sustain~\cite{D92}.  The threshold criterion for direct
linking as defined above represents an extreme limit where an agent ceases to
initiate new connections once he has two friends.  Nevertheless, it is still
possible that a popular person will make additional connections as a result
of being introduced to someone new.  It is worth emphasizing that the type of
transitive linking employed in this work plays an essential role in many
social networks, ranging from Granovetter's picture of the ``strength of weak
ties''~\cite{G73} to Facebook, where users are invited to link to the friends
of their Facebook friends~\cite{F,J08,TMP11}.

\begin{figure}[ht]
\centerline{\includegraphics[width=0.65\textwidth]{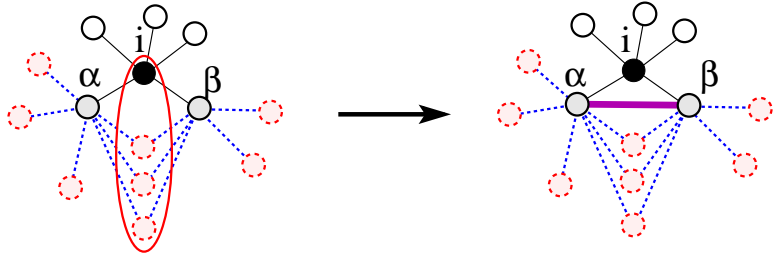}}
\caption{(color online) Illustration of threshold-controlled transitive
  linking.  An agent $i$ (solid) with five friends (open circles) is
  selected.  The selected agent introduces two of them, $\alpha$ and $\beta$
  (shaded).  These become friends (thick line) if the ratio of their mutual
  acquaintances (inside the oval) to the total acquaintances of either
  $\alpha$ or $\beta$ (dashed circles) exceeds a specified threshold.  Links
  outside this cluster are not shown. }
\label{model}
\end{figure}

The key feature of our acquaintance model is the imposition of distinct rates
for direct and transitive linking that are determined by the current state of
the network.  There are two different mechanisms that we implement to control
these rates:
\begin{enumerate}
\item \emph{Threshold-Controlled:} When two agents, $\alpha$ and $\beta$, are
  introduced by a common friend $i$, they connect if the number of their
  mutual friends $m_{\alpha\beta}$ (inside the oval in Fig.~\ref{model}) to
  the total number of friends of either agent (the degrees $k_\alpha$ and
  $k_\beta$) equals or exceeds a specified friendship threshold $F$.  That
  is, a connection occurs between $\alpha$ and $\beta$ if
  $m_{\alpha\beta}/d_{\alpha\beta}> F$, where $d_{\alpha\beta}=
  \mathrm{min}(k_\alpha,k_\beta)$.
\item \emph{Rate-Controlled:} The rates of transitive and direct linking are
  defined as $R$ and 1 respectively.
\end{enumerate}
The use of threshold-controlled transitive linking is motivated by the
observation that you are more likely to become friends with a newly introduced
person when the two of you already have many common friends.  The degree of
commonality can be a useful indicator how much two people have in common.

In each update step of our friendship model, an agent is selected at random.
If the degree of this agent equals 0 or 1, the agent links to another
randomly selected agent.  If the degree of the initial agent is 2 or larger,
a transitive link is created between two friends of the agent according to
the rates given above.  Notice that direct linking joins two clusters (here
clusters are defined as the maximal disconnected components of a graph),
while transitive linking merely ``fills in'' links within a cluster without
altering its size.  Updates continue until no more links can be created.
Also notice that once every agent has at least two friends, cluster mergings
no longer occur and links can be created only within a cluster.  In
threshold-controlled linking, the network reaches its final state when all
the friends of any agent are either linked to each other or can no longer
fulfill the threshold condition.  In rate-controlled linking, the final state
consists of a collection of complete subgraphs for any $R>0$; henceforth, we
term a complete subgraph of a network as a \emph{clique}.  The case $R=0$ is
unique, as will be discussed below.  In either case, a final state is reached
because geometrical constraints ultimately prevent the formation of
additional links.

\section{Threshold-Controlled Transitive Linking}

\begin{figure}[ht]
  \centerline{\subfigure[]{\includegraphics[width=0.10\textwidth]{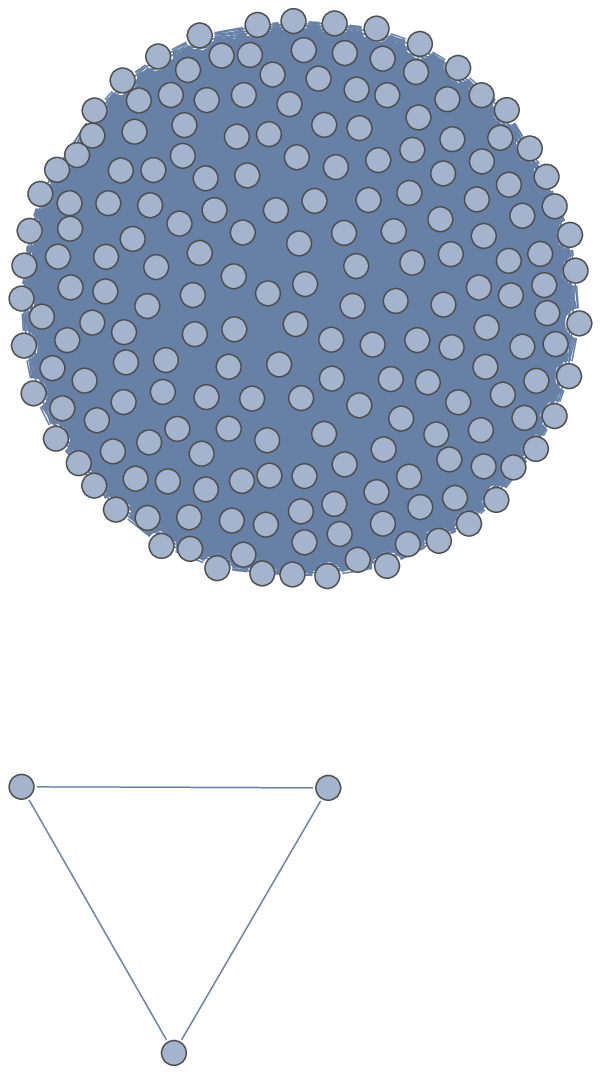}}\quad\subfigure[]{\includegraphics[width=0.25\textwidth]{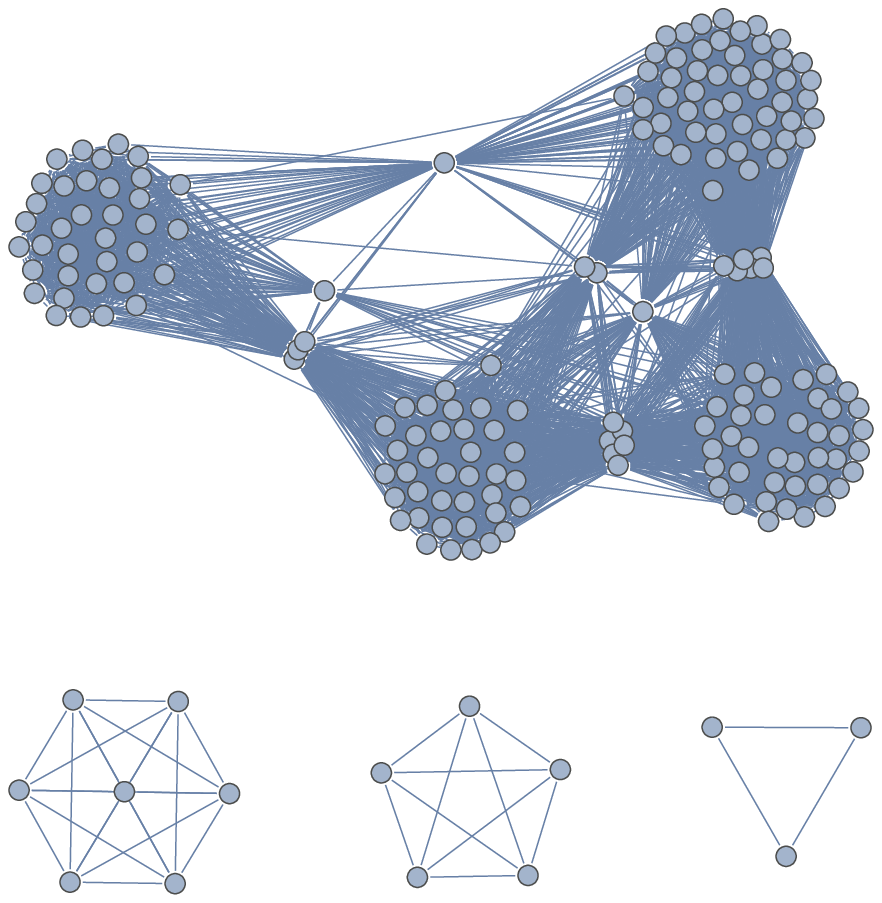}}\quad\subfigure[]{\includegraphics[width=0.28\textwidth]{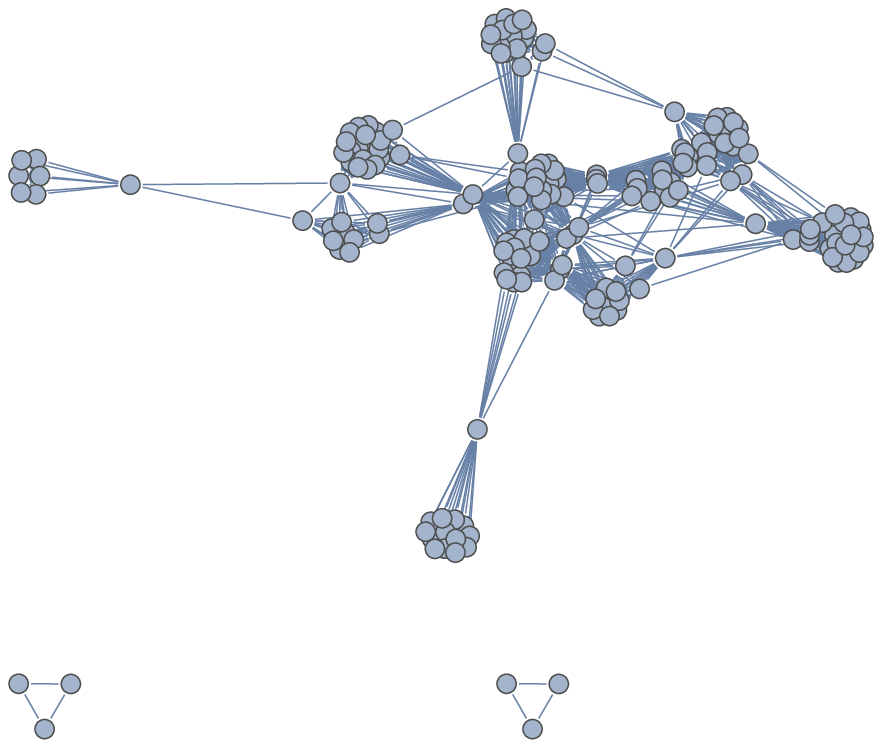}}\quad\subfigure[]{\includegraphics[width=0.28\textwidth]{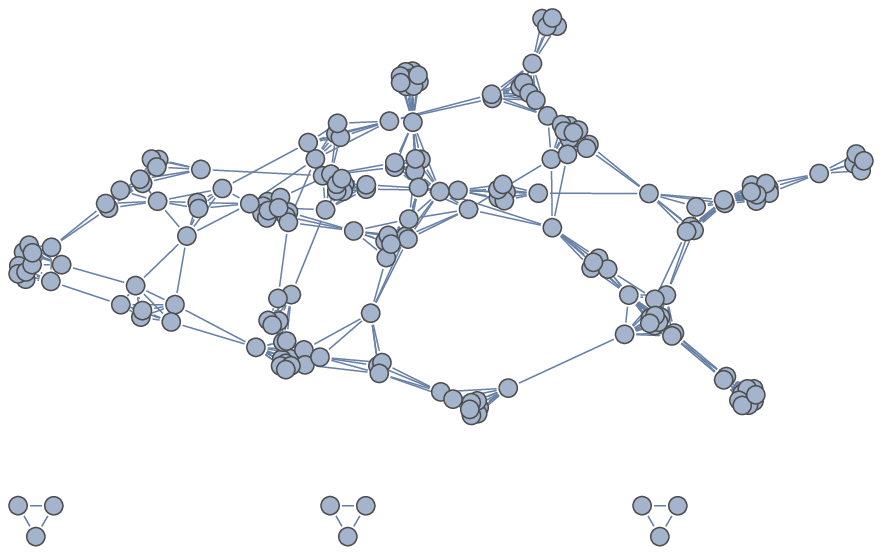}}}
  \caption{Typical networks of $N=200$ agents for threshold values: (a)
    $F=0.3$, (b) 0.35, (c) 0.4, and (d) 0.6.}
\label{snapshots}
\end{figure}

The most prominent feature of threshold-controlled transitive linking is the
emergence of highly-clustered substructures over a wide range of threshold
value (Fig.~\ref{snapshots}).

For small $F$, a new friendship is created nearly every time two individuals
are introduced by a mutual friend.  In this regime, the resulting graph is
nearly complete, but there also exists a small insular ``fringe'' population
that is comprised of small disjoint cliques (Fig.~\ref{snapshots}).  This
fringe arises because once every agent in a cluster has at least two links,
there is no mechanism for this cluster to merge with any other cluster.  Thus
even in the limiting case of $F=0$, the final state typically consists of
more than a single cluster, each of which is complete.  Concomitantly, the
largest cluster does not constitute the entire system for any value of $F$.
For intermediate values of $F$, the networks are highly clustered, as
illustrated in Figs.~\ref{snapshots}(b) and \ref{snapshots}(c), with the
largest cluster comprised of a small number of well-connected modules.

\begin{figure}[ht]
  \centerline{\subfigure[]{\includegraphics[width=0.45\textwidth]{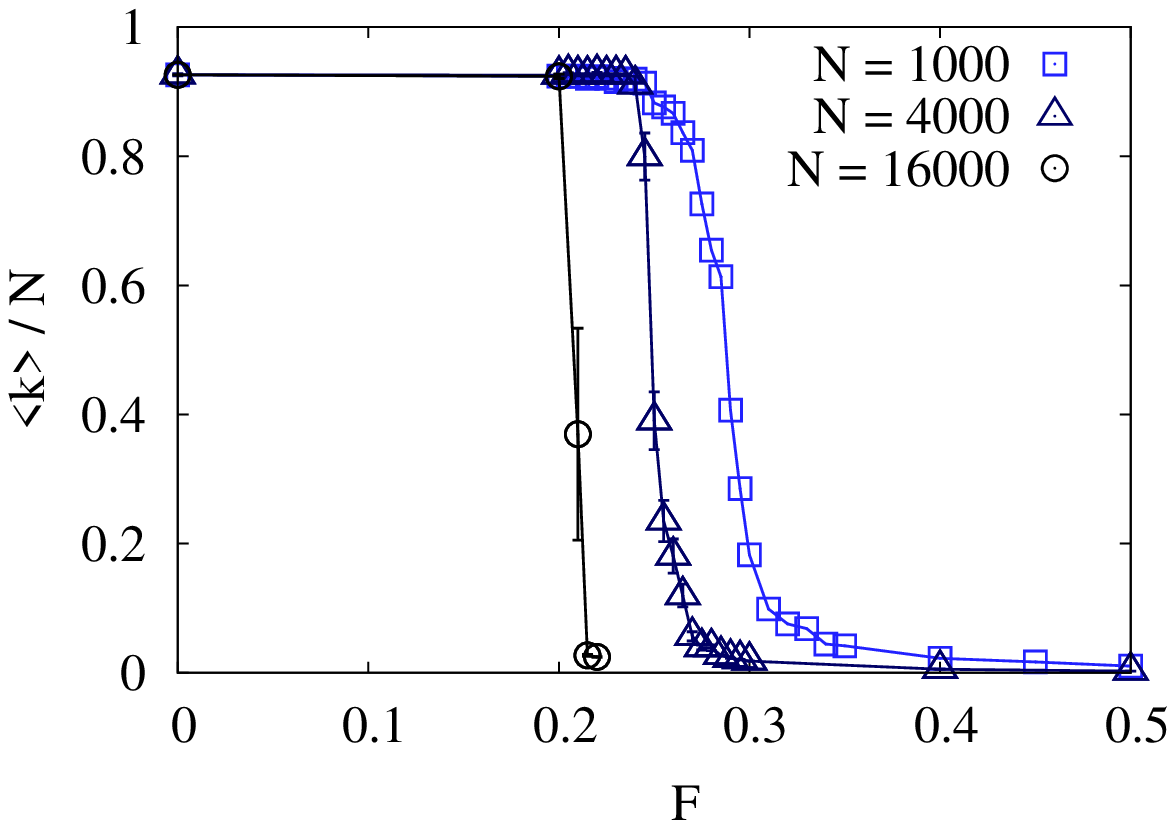}}\qquad\subfigure[]{\includegraphics[width=0.45\textwidth]{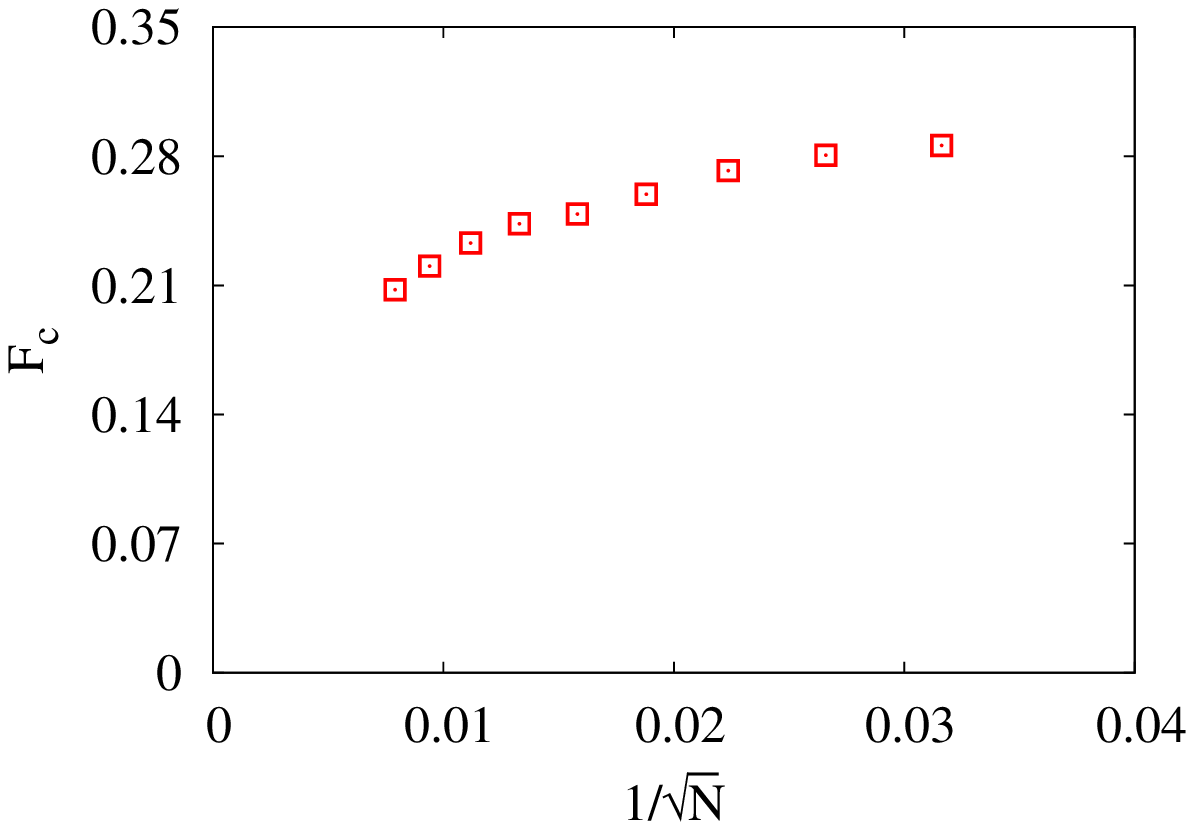}}}
  \caption{(a) Average degree divided by $N$ versus threshold. (b) Critical
    threshold versus $1/\sqrt{N}$.}
\label{transition-w-threshold}
\end{figure}

To help quantify the network and its clustering, we study the dependences of
the average degree and the distribution of community sizes as a function of
the threshold $F$.  The average degree exhibits a sharp change between a
dense and a sparse regime for $F$ in the range of 0.2--0.3 for network sizes
between $1000$ and $16000$ (Fig.~\ref{transition-w-threshold}(a)).  We define
the location of the transition as the point where $\langle k\rangle/N$, the
average degree divided by network size, equals $\frac{1}{2}$.  According to
this definition, the critical threshold value $F_c$ decreases very slowly
with $N$ (Fig.~\ref{transition-w-threshold}(b)).  From the data alone, it is
not evident whether a transition exists at non-zero $F_c$ for $N\to\infty$ or
whether this transition is a very slow finite-size effect.

A more direct way to understand how the network structure depends on $F$ is
by studying the community-size distribution.  Unlike real clusters, which are
unambiguously defined, the notion of community is somewhat fuzzy~\cite{FB07}.
Intuitively, a community is a community is a group of nodes that is densely
interconnected and is sparsely connected to nodes external to the community.
We adopt the definition given in Refs.~\cite{NG04,CNM04} in which the
communities of a given network are defined by the partition that maximizes
the modularity $Q$, defined by
\begin{equation}
\label{modularity-defn}
Q = \sum_{u v} \left[\left(\frac{A_{u v}}{2 L}\right) - \left(\frac{k_u}{2 L}\right)\left(\frac{k_v}{2 L}\right)\right]\delta\left(c_u, c_v\right)\,.
\end{equation} 
Here $A_{u v}$ is the adjacency matrix, with $A_{u v}=1$ if a link exists
between nodes $u$ and $v$ and $A_{u v}=0$ is no such link exists, $k_u, k_v$
are the degrees of these nodes, $L$ is the number of network links, and $c_u,
c_v$ label the communities that contain nodes $u$ and $v$.  The first term in
Eq.~\eqref{modularity-defn} is the fraction of links within all communities.
The second term gives the fractions of links that would exist within
communities if all connections were randomly rewired subject to the
constraint that all node degrees are preserved.  Thus the modularity is the
fraction of links within communities minus the fraction of links that would
exist within communities by chance.

\begin{figure}[ht]
  \centerline{\subfigure[]{\includegraphics[width=0.45\textwidth]{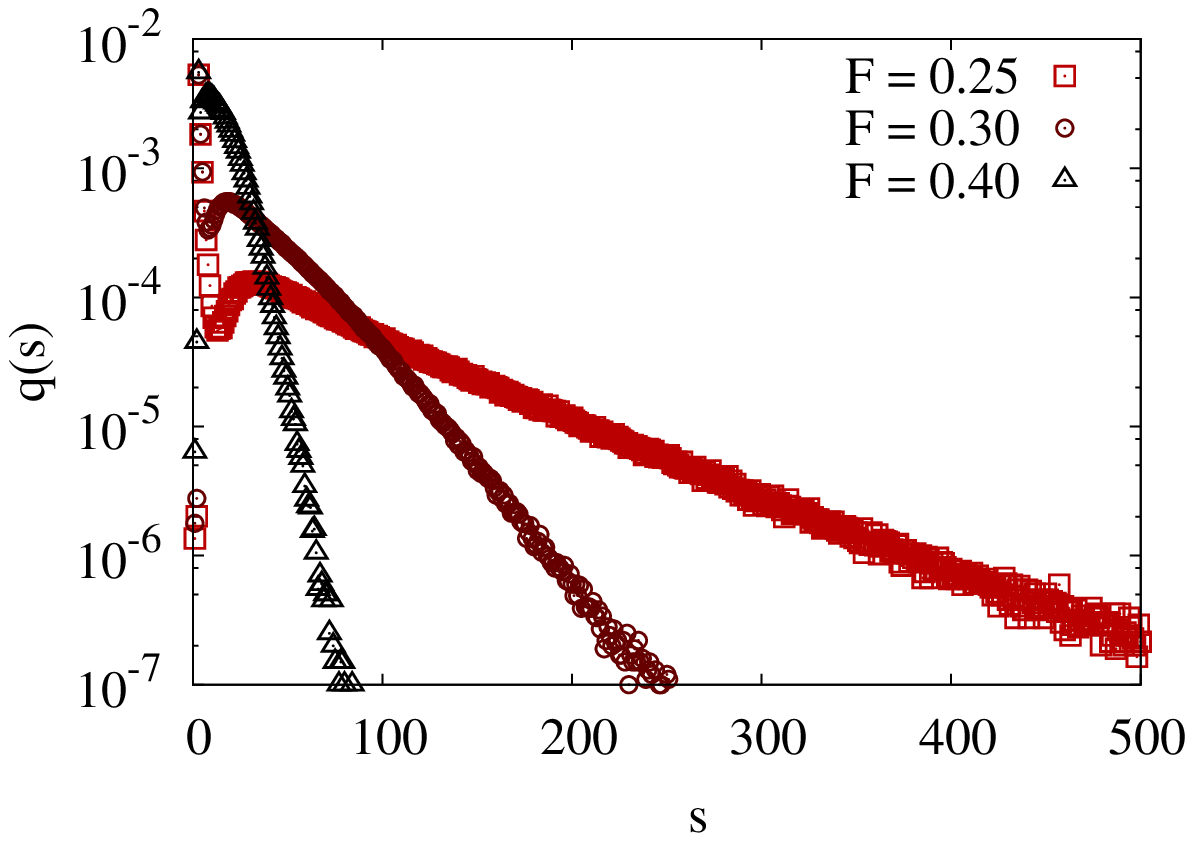}}\subfigure[]{\includegraphics[width=0.45\textwidth]{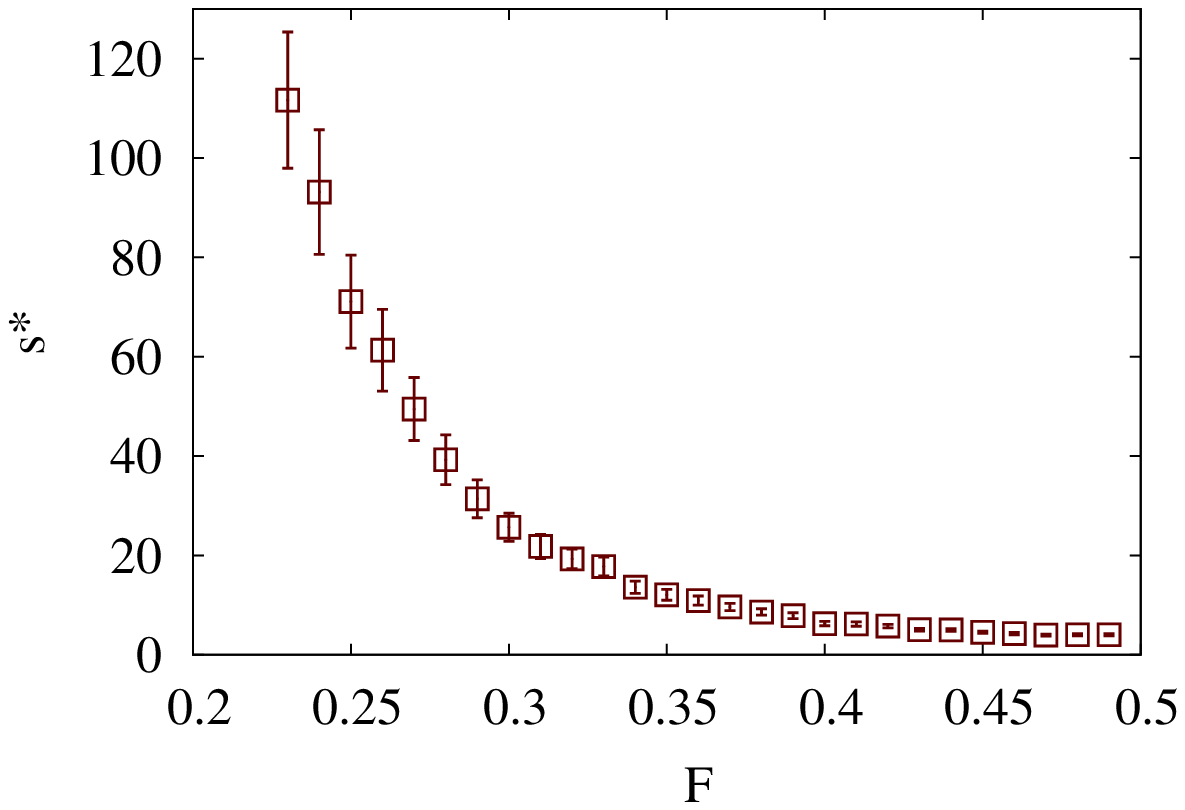}}}
  \caption{(a) Community size distributions for various thresholds $F$ for
    networks of $10^4$ nodes averaged over $10^4$ realizations each. (b) The
    slope of the exponential tail gives the characteristic community size.}
  \label{community-size-dist}
\end{figure}

To find the maximizing partition into communities, we apply the Monte Carlo
algorithm proposed by Blondel et al.~\cite{BGLF08}.  In this algorithm, the
network initially has the $N$ nodes in $N$ isolated communities.  Then for
each node $i$, we consider each of its neighbors $j$ and evaluate the gain of
modularity that would occur by placing node $i$ in the community of node $j$.
The node $i$ is ultimately included in the community for which this gain is
the largest (and positive).  Once this step of assigning individual nodes to
communities is done, the same fusion process is implemented on the current
network of communities.  This fusion of higher-order communities is repeated
until no further gain in modularity is possible.  At this point, the
algorithm gives the community size distribution, $q(s)$.  As shown in
Fig.~\ref{community-size-dist}(a), the tail of this distribution, averaged
over many realizations, decays as $e^{-s/s^*}$, with the characteristic
community size $s^*$ increasing as the threshold is decreased.  The data also
suggests that $s^*$ diverges as $F$ approaches $F_c$ from above
(Fig.~\ref{community-size-dist}(b)).  For $F<F_c$, communities are all
cliques, among which the largest has a size that scales linearly with $N$.
This apparent gelation phenomenon is best understood by investigating the
rate-controlled version of our friendship model, to be discussed in the next
section.

\section{Rate-Controlled Transitive Linking}

\begin{figure}[ht]
\centerline{ \subfigure[]{\includegraphics[width=0.375\textwidth]{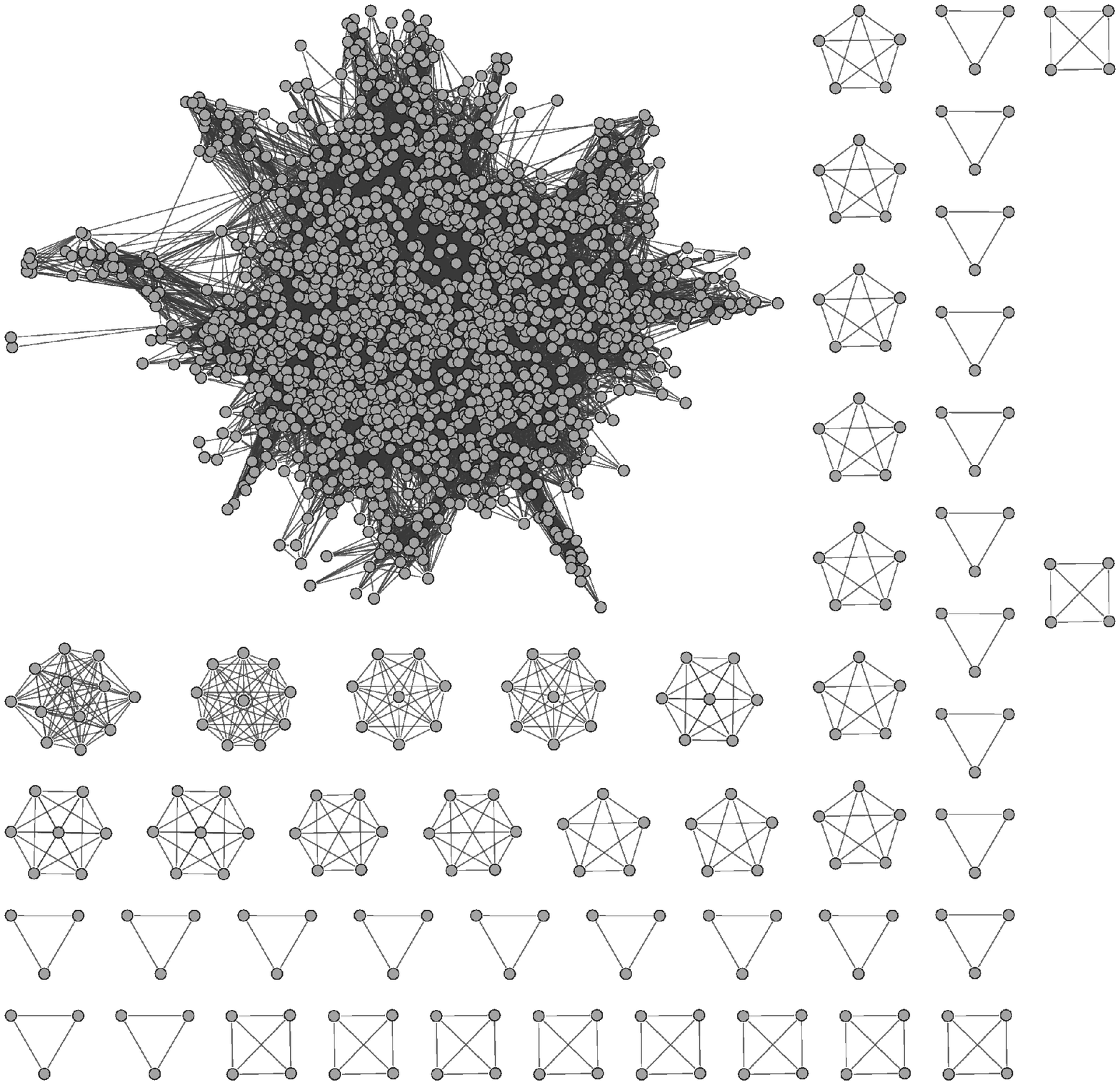}}\qquad\qquad\subfigure[]{\includegraphics[width=0.375\textwidth]{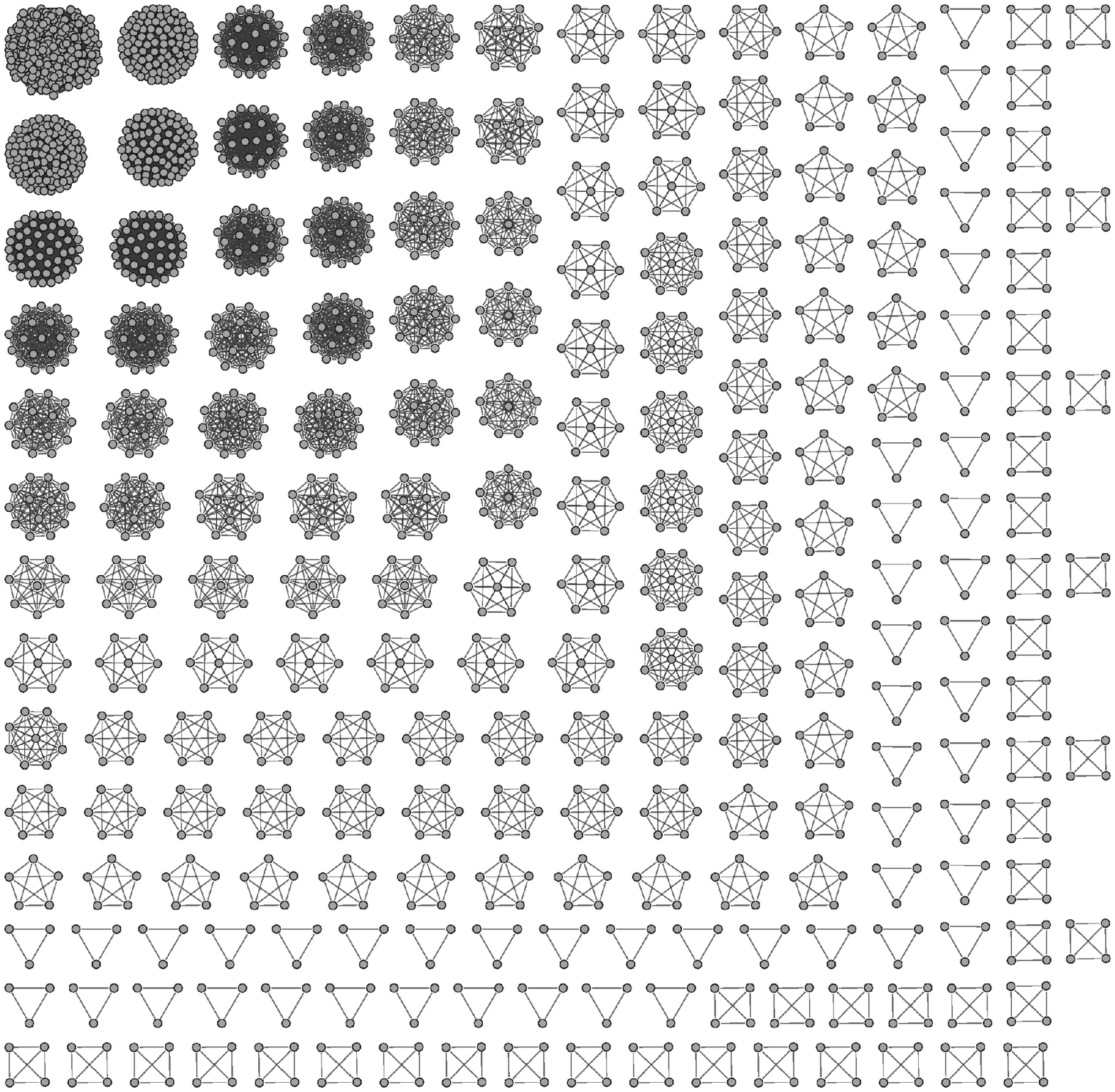}}}
\caption{Clusters at the instant when no nodes of degree 0 or 1 remain for a
  network of $2000$ nodes for: (a) $R=2$ (exponential
  cluster-size distribution) and (b) $R=15$ (power-law distribution).}
\label{final}
\end{figure}

While the threshold-controlled friendship model leads to networks with
visually striking community structure, many geometric and time-dependent
properties are more readily understood within the framework of
rate-controlled transitive linking.  As outlined in section~\ref{intro},
starting with an initial state of $N$ isolated nodes, nodes of degree 0 or 1
join to any other node in the network at rate 1, while a link between two
mutual friends of a node of degree 2 or greater occurs with rate $R$.  The
process ends when the network is partitioned into a set of cliques.
Figure~\ref{final} shows typical networks of $N=2000$ agents at the instant
when no nodes of degree 0 or 1 remain for: (a) $R=2$, where the cluster-size
distribution decays exponentially with size, and (b) $R\approx 15$, where the
distribution has a power-law decay.

\begin{figure}[ht]
\centerline{\subfigure[]{\includegraphics[width=0.375\textwidth]{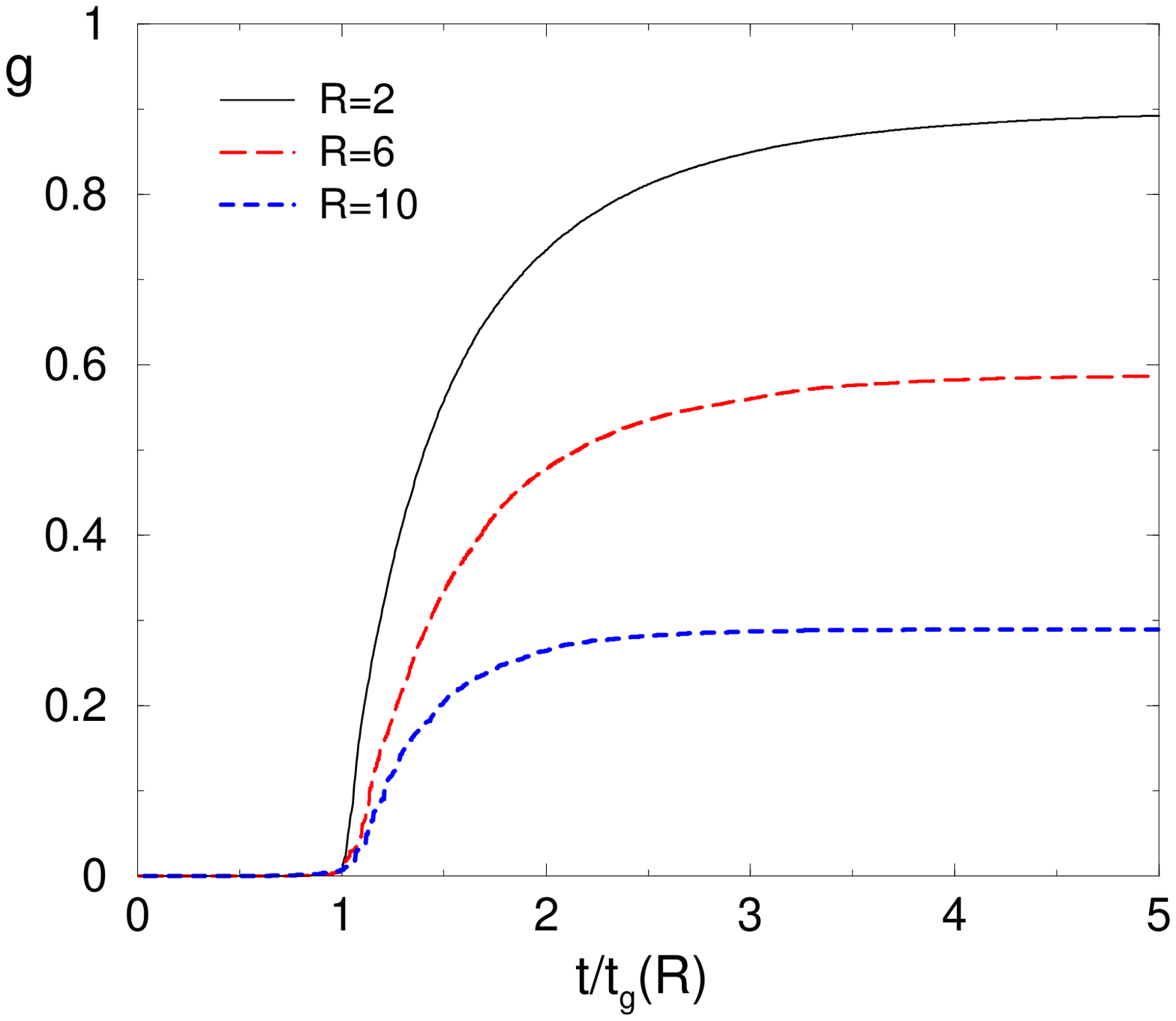}}\qquad\qquad\subfigure[]{\includegraphics[width=0.4\textwidth]{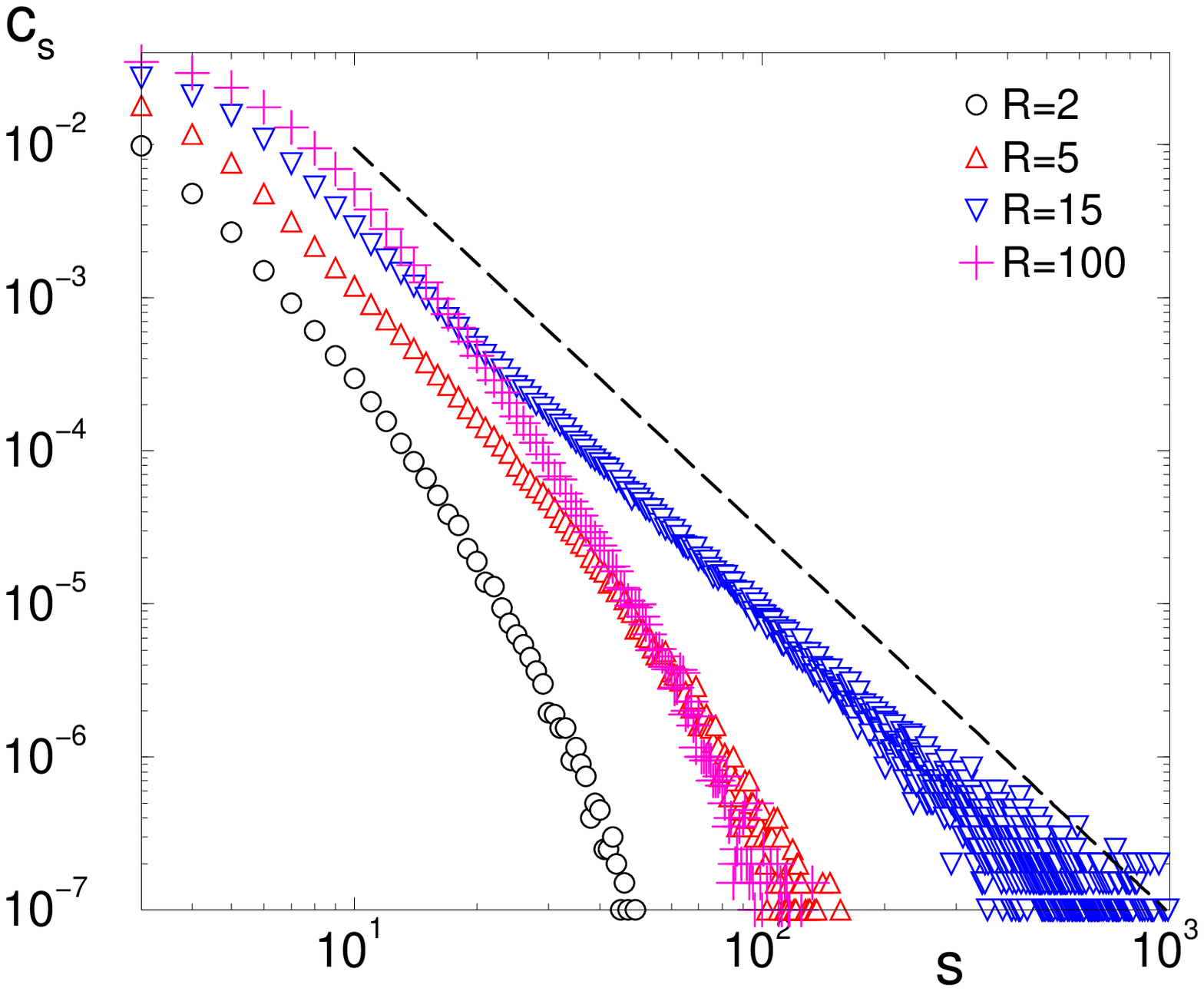}}}
\caption{Dependence of: (a) the fraction of agents in the largest cluster as
  a function of the normalized time $t/t_g(R)$ and (b) the cluster-size
  distribution $c_s$ versus size $s$ on a double logarithmic scale at
  infinite time, both for representative values of $R$.  Here $t_g(R)$
  is the gelation time for a given $R$.  In (b), linear behavior occurs only
  for $R=15$.  The dashed line has slope $-\frac{5}{2}$.}
\label{gel}
\end{figure}

In the range $R<R_c\approx 15$, rate-controlled transitive linking leads to
the emergence of a macroscopic cluster at a finite gelation time.  However,
this gelation phenomenon is \emph{incomplete}, because the fraction of agents
within this macroscopic cluster (also known as the gel fraction) saturates to
a value that is \emph{strictly} less than one as $t\to\infty$
(Fig.~\ref{gel}(a)).  Thus in addition to the single macroscopic cluster,
many small clusters persist forever.  This behavior strikingly contrasts with
classical gelation, where the gel encompasses the entire system when the
reaction runs to completion~\cite{L03,KRB10}.

The incompleteness of the gelation transition arises because the reaction is
controlled by \emph{active nodes}---those of degree 0 and degree 1.  When all
these nodes have been used up by linking to other nodes, there is no
possibility for additional cluster mergings.  All that can occur is
densification within each cluster.  Thus if multiple clusters happen to exist
when active nodes are exhausted, these clusters will persist forever.  In
spite of this incompleteness feature, the gelation transition itself seems to
conform to the classical mean-field description.  As $t$ approaches the
gelation time $t_g(R)$ from below, the concentration of clusters of size $s$,
$c_s$ gradually broadens and changes from an exponential decay as a function
of $s$ to an algebraic decay (Fig.~\ref{gel}(b)).  At the gelation time,
$c_s\sim s^{-\alpha}$, with $\alpha\approx \frac{5}{2}$, as in classical
gelation.  The gelation time itself diverges for $R\ge R_c \approx 15$.  In
the regime where $R>R_c$, transitive linking events quickly use up all active
nodes, which are the catalysts for cluster merging.  Because the average
cluster size is still small at the instant when active nodes are used up,
gelation is suppressed for $R>R_c$.  In this non-gelling regime, the
cluster-size distribution decays exponentially with size at all times.

\begin{figure}[ht]
  \centerline{\includegraphics[width=0.45\textwidth]{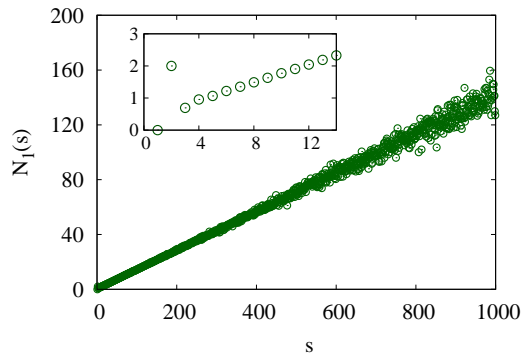}}
  \caption{ Average number of active nodes within clusters of size $s$ at
    fixed time $t=0.98$ for network of $10^6$ nodes and for the case of
    $R=2$.  The data represents an average over $100$ realizations.  The
    inset shows the small-$s$ behavior.}
  \label{n1}
\end{figure}

We can make a more quantitative correspondence between rate-controlled
transitive linking and gelation by mapping the former onto a version of
classical product kernel aggregation~\cite{L03,KRB10}.  This correspondence
relies on our observation from simulations that the number of active nodes in
a cluster of size $s$ at any given time is proportional to $s$
(Fig.~\ref{n1}).  The proportionality constant is time dependent because the
concentration of nodes of degree one (which we term leaf nodes) nodes change
with time. As shown in the inset to Fig.~\ref{n1}, the contribution of
clusters of size 1 and 2 deviates from the overall linear trend in the main
figure.  In particular, a monomer (a node of degree 0) has no leaf nodes
while a dimer always has two leaf nodes (each of degree 1).  For clusters of
size $s>2$, we assume that the average number of leaf nodes is given by
$N_1(s,t) = \lambda(t)\,s$.  Then the total number of leaf nodes, $n_1$, can
be expressed through the cluster-size distribution: $n_1 = 0\cdot c_1 + 2 c_2
+ \sum_{k\geq 3}\lambda\, k\, c_k$. Combining this equation with the
conservation law $\sum_{k\geq 1} k c_k=1$ we get $n_1= 2 c_2 + \lambda (1 -
c_1 - 2 c_2)$, from which
\begin{equation}
  \lambda(t) = \frac{n_1  - 2 c_2}{1 - c_1 - 2 c_2}
\end{equation}

\begin{figure}[ht]
\centerline{\includegraphics[width=0.5\textwidth]{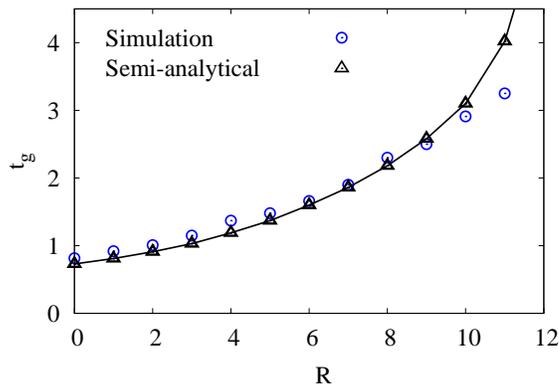}}
\caption{Comparison of gelation times, $t_g$ obtained by semi-analytical
  calculation of $M_2$ and $t_g$ obtained directly from simulations}
\label{compare-tg}
\end{figure}

We now now write the following product-kernel-like aggregation equations for
the cluster-size distribution \cite{L03,KRB10}, in which we separately
account for the evolution of monomers and dimers:
\begin{equation}
\label{general-R-c1}
\begin{split}
\dot{c}_1   &= - c_1 - (c_1 + n_1)c_1\,,\\
\dot{c}_2   &= c_1^2 - 2 c_2 - 2 c_1 c_2 - 2 n_1 c_2\,,\\
\dot{c}_k   &= c_1 \big[c_{k-1}(k-1) - k c_k\big] + 2 c_2 \big[(k-2)c_{k-2}-k c_k\big]\\
                   &~~~+ \sum_{j\ge 3} \lambda \, j\, (k-j)\, c_j c_{k-j} -  \sum_{j\ge 3}
\lambda\, j \,k \,c_j c_k - \lambda\, k\, c_k\,,  \qquad k\geq 3,
\end{split}
\end{equation}
where the dot denotes the time derivative. From these equations, the second
moment of the cluster-size distribution evolves according to
\begin{equation}
\label{general-R-M2}
\begin{split}
\dot{M}_2 &=16 c_2^2 + 2 c_1 c_2 - n_1 (c_1 + 8 c_2) + (2 c_1 + 8 c_2)M_2(1-\lambda)\\
&~~~ + \lambda(2 M_2^2 + c_1 + 8 c_2 - c_1^2 - 16 c_2^2 - 10 c_1 c_2)\,.
\end{split}
\end{equation}
If gelation does occur, the second moment $M_2$ would diverge at a gelation
time $t_g(R)$.  However, we are unable to write a closed equation for the
concentration of leaf nodes $n_1$ and thereby solve for $M_2$ and $t_g$.
Thus to find $t_g$, we take the value $n_1$ from simulations and use it solve
the equations for $\dot{c}_1$, $\dot{c}_2$ and $\dot{M}_2$ numerically.  This
approach gives good agreement with the value of $t_g(R)$ obtained by direct
simulations (Fig.~\ref{compare-tg}), as long as $R$ is not close to $R_c$.
We can also find $R_c$ from our semi-numerical method by scanning across
different values of $R$ and finding the value of $R$ where $M_2(R,t=\infty) =
\infty$ and $M_2(R+\Delta R,t=\infty)<\infty$.  This approach gives a lower
value for $R_c \approx 12.1$ compared to $R_c \approx 15$ directly from
simulations and thus gives a sense of the accuracy of our semi-numerical
method.

\section{Extremal Limits of Transitive Linking}

To develop additional insights, we now investigate our friendship model in
the extremal limits of no transitive linking or infinitely rapid transitive
linking.  The former case may be achieved in the threshold model with
$F=\infty$ or in the rate model with $R=0$.  The latter is achieved in the
rate model by setting $R=\infty$.  In these limiting cases we can obtain
useful insights about some basic network properties by analytical means.

\subsection{No Transitive Linking}

\begin{figure}[ht]
\centerline{\includegraphics[width=0.6\textwidth]{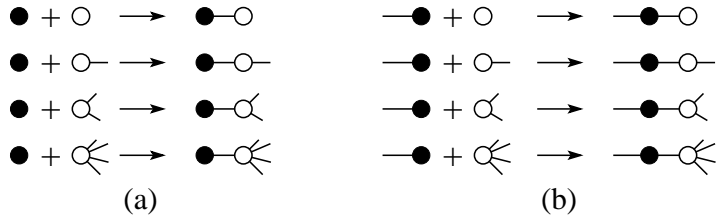}}
\caption{The elemental evolution steps without transitive linking: (a)
  reaction channels for a node of degree 0 and (b) for a node of degree 1.
  The solid circle denotes the initial node. }
\label{R=0}
\end{figure}

In the absence of transitive linking, the network evolves by a constrained
aggregation process that is mediated only by active nodes---those of degrees
0 or 1 (Fig.~\ref{R=0}).  The network stops evolving when these active nodes
no longer exist.  By enumerating all the possible ways that an active node
can interact (Fig.~\ref{R=0}), the concentrations $n_k$ of agents of degree
$k$ evolve according to
\begin{equation}
\label{ME-0}
\begin{split}
\dot n_0 &= -n_0 (1+a) \\
\dot n_1 &=(n_0-n_1)(1+a) \\
\dot n_2 &= n_1 +(n_1-n_2)a \\
\dot n_k &= (n_{k-1}-n_k) a, \qquad k\geq 3\,.
\end{split}
\end{equation}
Here $a=n_0+n_1$ is the concentration of active agents.  One can verify that the conservation law $\sum_{k\geq 0}\dot n_k=0$ is obeyed and that  the mean degree grows according to 
$\frac{d}{dt}\,\langle k\rangle =\sum_{k\geq 0}k\dot n_k=2a$.

The rate equations \eqref{ME-0} admit an exact, albeit implicit, solution.  To
simplify the first two lines of Eqs.~\eqref{ME-0}, we introduce the time-like
variable $dT = dt(1+a)$ to recast these equations as $n_0' = -n_0$ and $n_1'
= n_0-n_1$, where the prime denotes differentiation with respect to $T$.  The
solution is
\begin{equation}
\label{n01:T}
n_0 = e^{-T}\,, \qquad\qquad n_1 = Te^{-T}\,.
\end{equation}
Consequently the original and modified time variables are related by
\begin{equation}
\label{tT}
t = \int_0^T \frac{dT'}{1+(1+T')\,e^{-T'}}
\end{equation}
Equations \eqref{n01:T}--\eqref{tT} provide the exact, but implicit solution
for the densities of agents with degree 0 and degree 1.

To obtain more explicit results, we need to relate $t$ and $T$.  To this end,
we write
\begin{align}
\label{T-t}
T-t &= \int_0^T dx\left[1- \frac{1}{1+(1+x)e^{-x}}\right]\nonumber \\
& =  \int_0^\infty dx \left[\frac{1+x}{1+x+e^x}\right]- 
\int_T^\infty dx \left[\frac{1+x}{1+x+e^x}\right]\nonumber\\
&\equiv \alpha -\mathcal{O}(T\, e^{-T}) = \alpha-\mathcal{O}(t\, e^{-t})\,,
\end{align}
with the  value of $\alpha$ determined numerically to be $1.2802837\ldots$.
Thus the densities of active agents asymptotically vary as
\begin{equation}
n_0\to e^{-\alpha}\, e^{-t}, \qquad \qquad n_1\to e^{-\alpha}\, t \,e^{-t}\,.
\end{equation}
We estimate the time when active agents are exhausted by the criterion
$n_1(t^*)=1/N$; namely, a single node of degree 1 remains in a network of $N$
nodes at the completion time $t^*$.  From the above asymptotic dependence of
$n_1$, the completion time is given by $t^*\simeq \ln N +\ln\ln N$.

To determine the density of agents with two or more friends, we define a
second time-like variable $d\tau=a\,dt$ to recast the last line of
Eq.~\eqref{ME-0} as
\begin{equation}
\label{nkT}
\frac{dn_k}{d\tau} = n_{k-1} - n_k\qquad k\geq 3\,.
\end{equation}
Assuming that we know $n_2$, we use the Laplace transform method to solve
\eqref{nkT} and then invert the Laplace transform to give the recursive
solution
\begin{equation}
\label{nk_sol}
n_k(\tau)=\frac{1}{(k\!-\!3)!}\int_0^\tau d\tau'\, (\tau\!-\!\tau')^{k-3} 
e^{\tau'-\tau}\,n_2(\tau')
\end{equation}
for $k\geq 3$. To determine $n_2$ we rewrite its evolution equation, the
third line of \eqref{ME-0}, as
\begin{equation}
\label{n2T}
\frac{dn_2}{dT} + n_2\,\frac{d\tau}{dT} = n_1\,.
\end{equation}
Integrating this equation and making use of $\tau=T-t$ gives the implicit
solution
\begin{equation}
\label{n2_sol}
n_2(T) = e^{-\tau}\int_0^T dT'\, T' e^{-t(T')}\,.
\end{equation}
In the $t\to\infty$ limit, the density of nodes of degree 2 is
\begin{eqnarray*}
n_2(\infty)= e^{-\alpha} \int_0^\infty dT\, T\,\exp\left[-\int_0^T \frac{dT'}{1+(1+T')e^{-T'}}\right]=0.6018583\ldots
\end{eqnarray*}

We now exploit this result to determine the density of nodes of arbitrary
degree in the limit $t\to\infty$.  First, notice that, by definition, the
rescaled time $\tau=T-t$.  Thus from Eq.~\eqref{T-t}, $\tau\to\alpha$ as
$t\to\infty$, so that the fraction of nodes of degree $k>2$ at infinite time
is given by
\begin{equation}
\label{nk_final}
n_k(\infty) = \frac{1}{(k-3)!}\int_0^\alpha d\tau\,  (\alpha-\tau)^{k-3} e^{\tau-\alpha}\,n_2(\tau)\,.
\end{equation}
The large-$k$ asymptotic is simpler to determine, since the
small $\tau$ limit of $n_2(\tau)$ makes the dominant contribution to the
above integral.  For early times, one gets $\tau\simeq
t$ and $n_2\simeq 2t^2$ as $t\to 0$.  Substituting $n_2(\tau)\simeq 2\tau^2$
into \eqref{nk_final} gives, for $k\gg 1$,
\begin{equation}
\label{nk_infinite}
n_k(\infty) \simeq 4\,e^{-\alpha}\,\,\frac{\alpha^k}{k!}~.
\end{equation}

We may similarly obtain the average degree of the final network.  We first
rewrite the evolution equation for the mean degree as $\frac{d\langle
  k\rangle}{d\tau}=2$, so that $\langle k\rangle = 2\tau$.  The mean degree
therefore starts at zero and increases to $2\alpha=2.560567483\ldots$ as
$t\to\infty$.  Thus when transitive linking is not allowed, the final network
is sparse and only slightly more dense than a tree. [For a tree of $N$ nodes, 
the average degree is $2(1-\tfrac{1}{N})$.]

\subsection{Infinite Transitive Linking Rate}

\begin{figure}[ht]
\centerline{\includegraphics[width=0.6\textwidth]{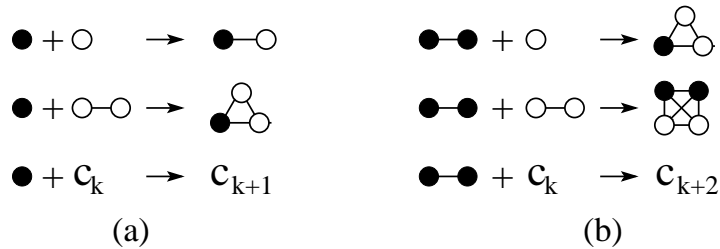}}
\caption{Elemental processes when transitive linking occurs with infinite
  rate for a node of degree: (a) 0 and (b) 1.  The initial cluster is shown
  solid. }
\label{R=inf}
\end{figure}

The complementary situation where transitive linking is infinitely rapid is
also tractable because each cluster is always a clique and may thus be fully
characterized by its size.  The update steps for this limiting case are
summarized by (see Fig.~\ref{R=inf}):
\begin{enumerate}
\item[(i)] Select an active agent that connects to another agent in a cluster
  of size $k$.  If the initial agent has degree 0, the cluster size increases
  from $k$ to $k+1$, If the initial agent has degree 1, the cluster size
  increases to $k+2$.
\item[(ii)] After each growth event, all possible links within the enlarged
  cluster are immediately filled in so that the resulting cluster remains
  complete.
\end{enumerate}
Once all agents of degree 0 or degree 1 are used up, the network has reached
a final state that consists of a collection of cliques.  

\begin{figure}[ht]
\centerline{\includegraphics[width=0.45\textwidth]{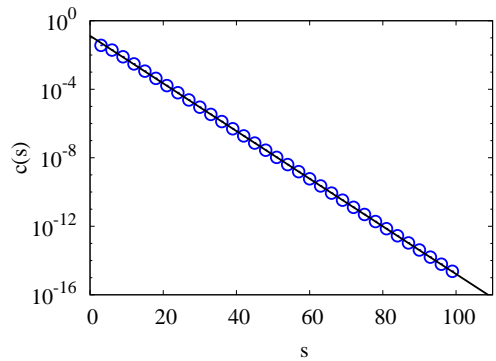}}
\caption{Numerical solution to Eqs.~\eqref{ME-inf} (circles) and an
  exponential fit to this data (line).}
\label{R=inf-clustersizedist}
\end{figure}

Using a rate equation approach, we can determine basic properties of this
final clique-size distribution.  From the reaction steps outline above, the
concentrations of clusters of size $k$, $c_k$, evolve according to
\begin{equation}
\label{ME-inf}
\begin{split}
\dot c_1 &= - c_1- c_1^2 - 2 c_1 c_2\  \\
\dot c_2 &=c_1^2 -2c_1c_2 -2c_2-4c_2^2  \\
\dot c_k &=−c_1\big[(k\!-\!1)c_{k-1}\!-\!kc_k\big]  
+2c_2\big[(k\!-\!2)c_{k-2}\!-\!kc_k\big],\qquad k\geq 2\,.
\end{split}
\end{equation}
Let us first determine the concentration of active clusters---monomers of
size 1 or dimers of size 2.  Keeping the two largest terms in equations for
$\dot c_1$ and $\dot c_2$, one finds the following long-time behaviors for the monomers and dimer
concentrations: 
\begin{equation}
c_1\simeq (2e^t-1)^{-1}\to \tfrac{1}{2}\,e^{-t}, \qquad c_2\simeq \tfrac{t}{4}\, e^{-2t}\,.
\end{equation}
As one might anticipate, the concentrations $c_1$ and $c_2$ asymptotically
decay exponentially with time.  Since these are the catalysts for the
reactions of larger clusters, the network quickly reaches a final static
state.  By numerically integrating the master equations \eqref{ME-inf}, the
cluster-size distribution evolves to a final, time-independent form $c_s\sim
e^{-s/s^*}$ for large $s$, with $s^*\approx 3.21$
(Fig.~\ref{R=inf-clustersizedist}.).

From the rate equations, the first three integer moments of the cluster-size
distribution, $M_n\equiv\sum k^n c_k$, evolve according to
\begin{align*}
\dot M_0=-c_1-2c_2,\qquad\qquad  \dot M_1=0, \qquad \qquad \dot M_2= 2M_2(c_1+4c_2)\,.
\end{align*}  
Using the above asymptotic behaviors of $c_1$ and $c_2$, we see that the
cluster density $M_0$ approaches a non-zero asymptotic value exponentially
quickly in time.  Numerically, the final concentration of clusters is given by
$M_0(\infty) = 0.1666474164\ldots$.  Similarly, the average cluster size
saturates to a finite value as $t\to\infty$. 

\section{Facebook Networks and the ``Soft Cutoff'' Model}

We now investigate whether our acquaintance model can account for the
observed features of real networks.  Useful empirical datasets with which we
can make such a comparison are anonymized Facebook networks from 100
well-known universities in the US~\footnote{The anonymized data was kindly
  provided by Mason Porter.}.  Many of these networks exhibit significant
clustering, although not to the same degree as in our acquaintance model.
One possible reason for the stronger clustering in our model is the sharp
cutoff between direct and transitive linking.  In our model, once an agent
has acquired two friends, he no longer can make additional friends directly,
a social interaction that would join two communities.  Thus the sharp cutoff
between direct and transitive linking enhances insularity.  However in real
social interactions, individuals with many friends can still engage in direct
linking, a mechanism that decreases the modularity and the clustering
coefficient of the resulting network compared to our idealized model with a
sharp cutoff.

\begin{figure}[ht]
\centerline{\subfigure[]{\includegraphics[height=2.0in,width=0.3\textwidth]{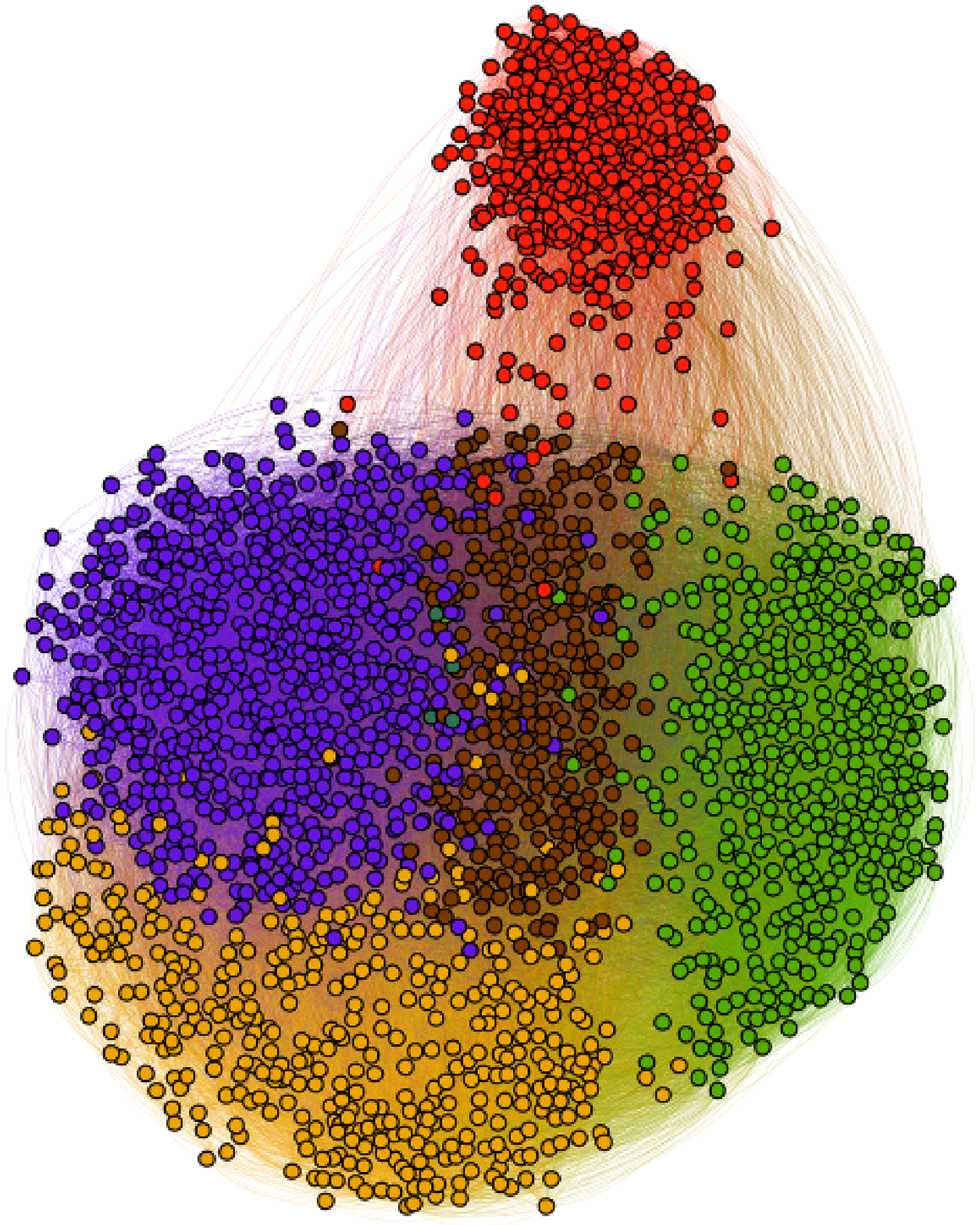}}\qquad\qquad\qquad
\subfigure[]{\includegraphics[width=0.3\textwidth]{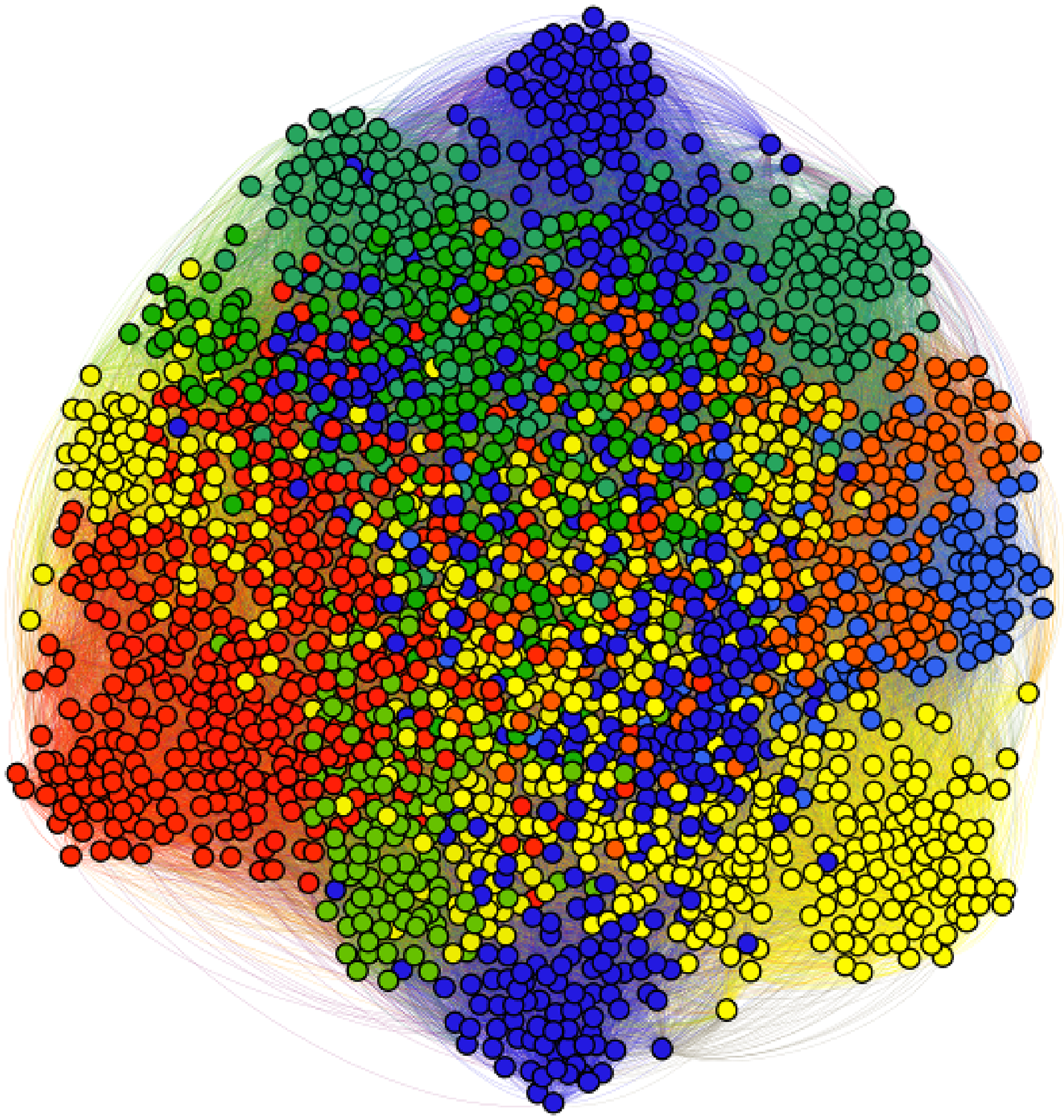}}}
\caption{Comparison between (a) the Facebook network of a well-known US
  university, with clustering coefficient $C=0.29$ and modularity $Q=0.45$,
  and (b) a realization of our soft-cutoff acquaintance model of identical size
  to (a) in which the rate $\lambda$ that an agent makes a direct link is an
  exponentially decaying function of its degree, $\lambda= e^{-k/k^*}$.  In
  (b), the choice $k^* = 7.3$, $F = 0.17$ leads to $C=0.32$ and $Q=0.45$.
  Communities are indicated by the different colors.}
\label{comparison}
\end{figure}

We are therefore led to define a ``soft cutoff'' model, in which the rate
$\lambda$ at which an agent engages in direct linking is a decreasing
function of number of his friends, $\lambda =e^{-k/k^*}$.  We argue that this
soft cutoff more realistically accounts for how an individual makes new
friends.   Figure~\ref{comparison} shows a comparison between an anonymized
Facebook network with $N=2252$ nodes and $L=84387$ links from a well-known US
university and a realization of our soft-cutoff acquaintance model with the
same number of nodes and links.  For the latter, the values of $k^*=7.3$ and
$F=0.17$ give modularity and clustering coefficients close to those of the
Facebook example.

\begin{figure}[ht]
\centerline{\subfigure[]{\includegraphics[width=0.32\textwidth]{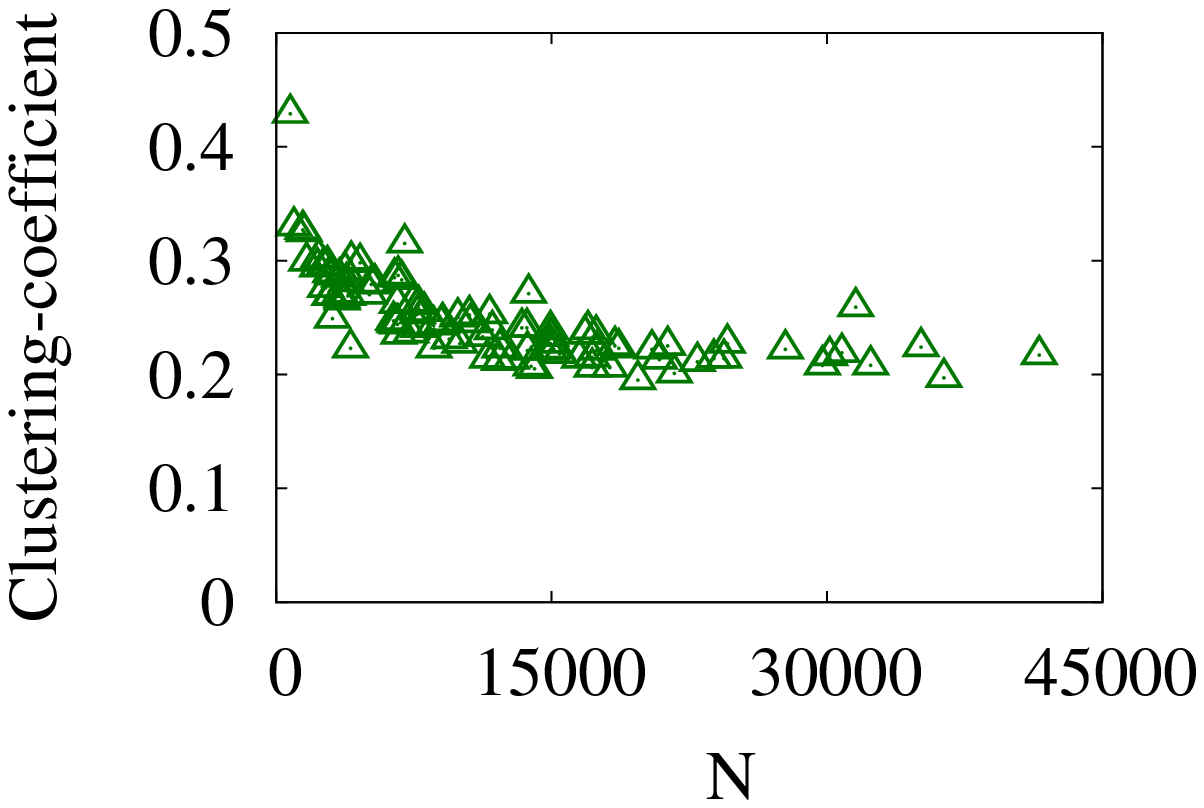}}
  \quad 
\subfigure[]{\includegraphics[width=0.32\textwidth]{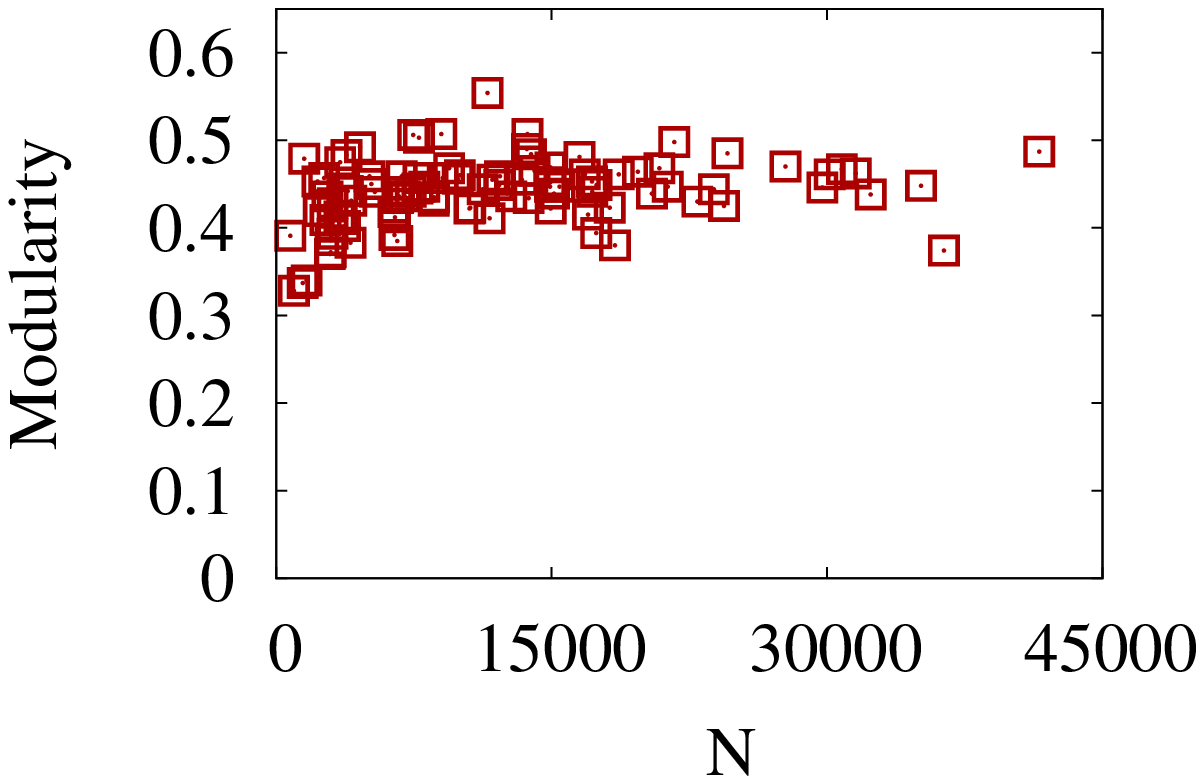}}\quad
\subfigure[]{\includegraphics[width=0.32\textwidth]{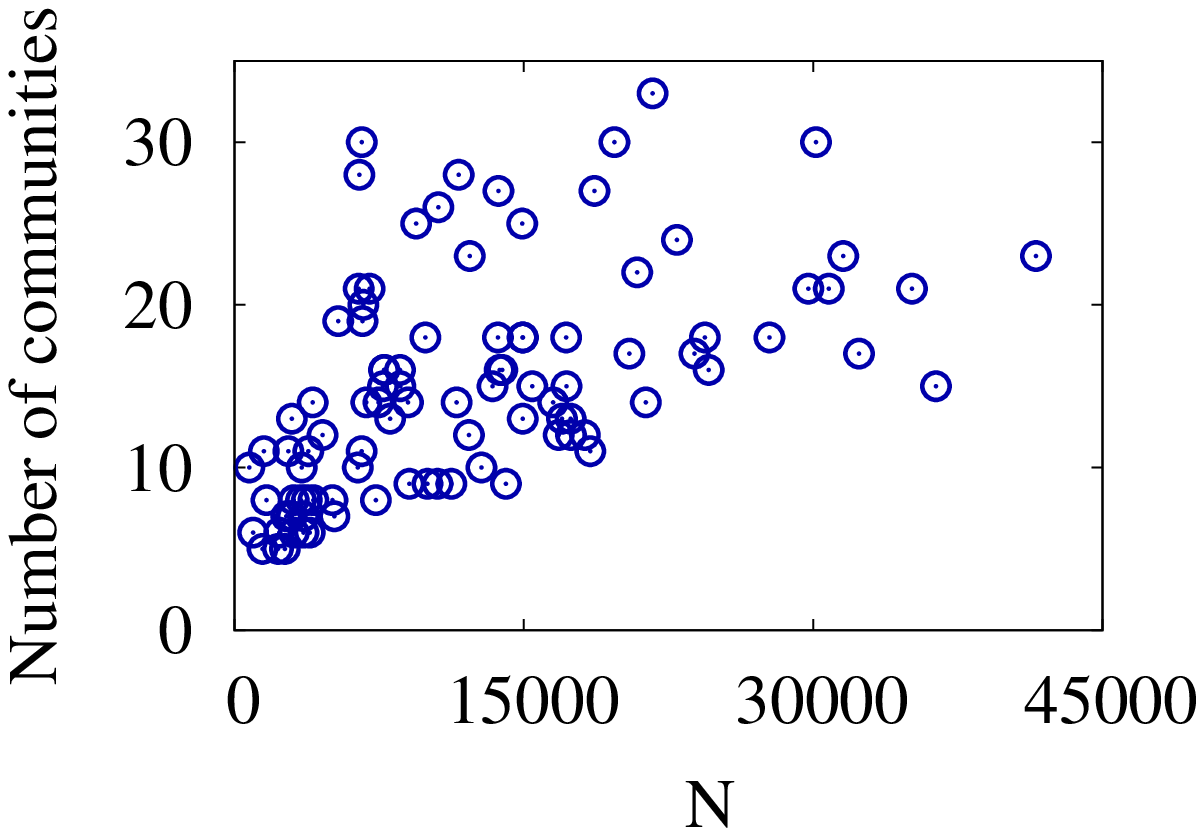}}}
\caption{(a) Clustering coefficient $C$, (b) modularity $Q$, and (c) number
  of communities for the Facebook networks of 100 well-known US
  universities.}
\label{facebook-data}
\end{figure} 

\begin{figure}[ht]
\centerline{\subfigure[]{\includegraphics[width=0.32\textwidth]{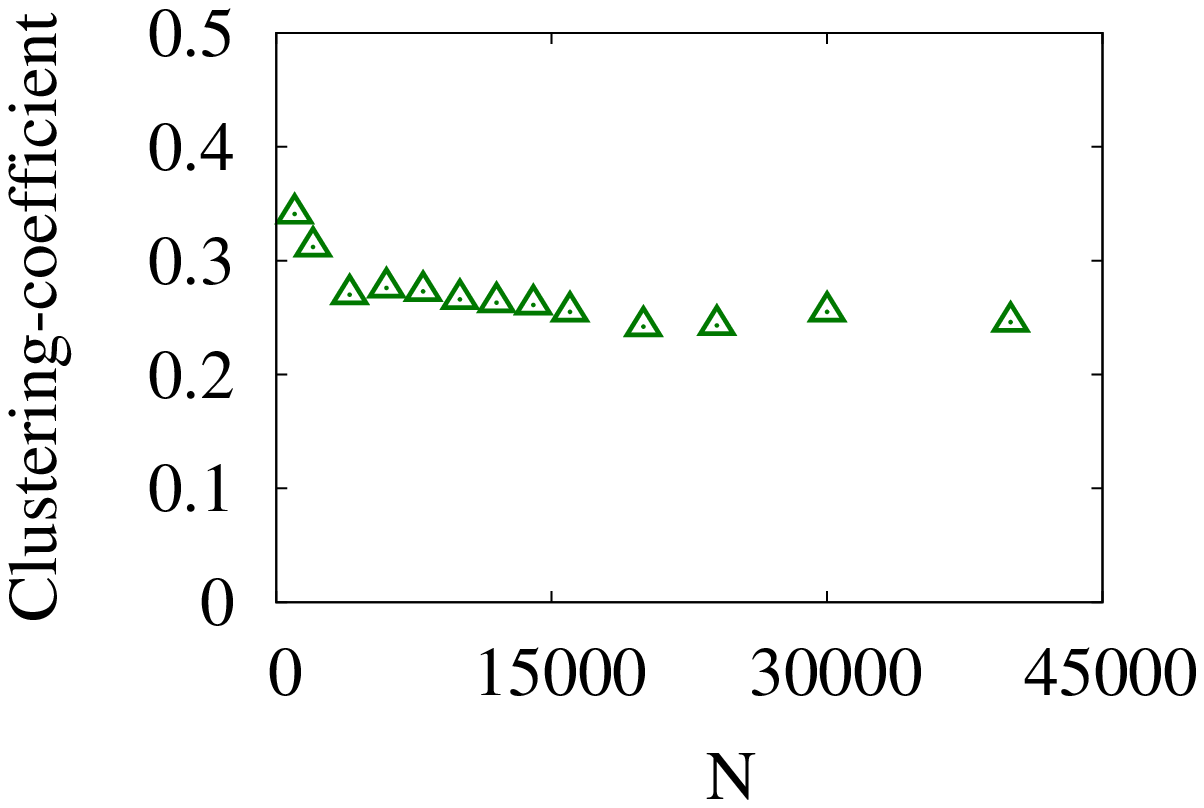}}\quad
\subfigure[]{\includegraphics[width=0.32\textwidth]{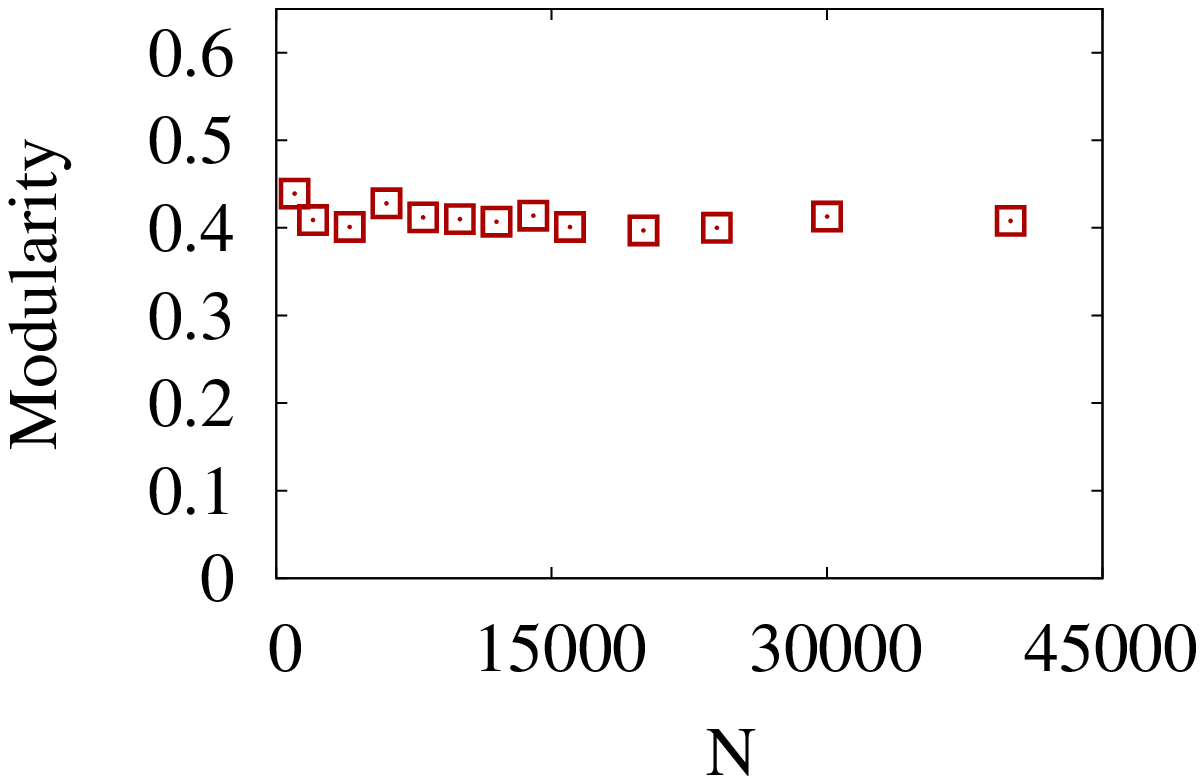}}\quad
\subfigure[]{\includegraphics[width=0.32\textwidth]{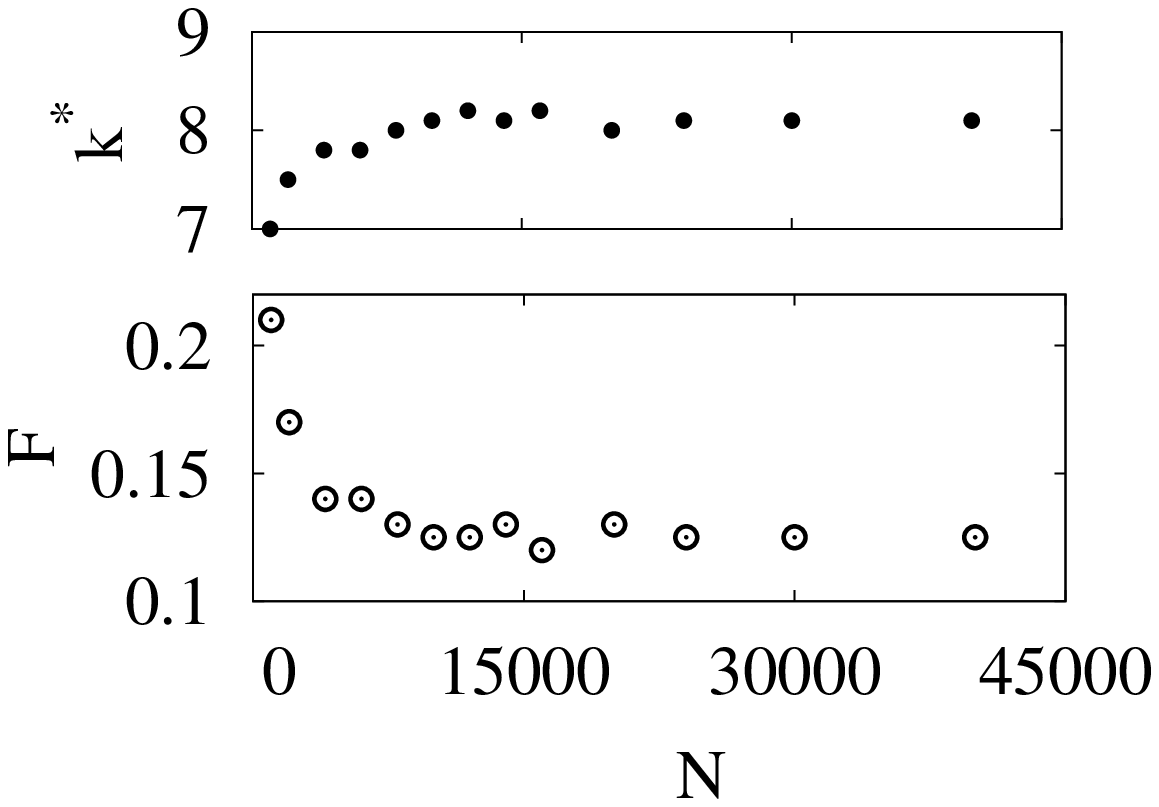}}}
\caption{(a) Clustering coefficient and (b) modularity from the soft-cutoff
  model. (c) The corresponding parameters $k^*$ and $F$ in the soft-cutoff model.}
\label{model-data}
\end{figure} 

This ability to match our soft-cutoff acquaintance model to Facebook networks
extends to examples with $N$ ranging from $1000$ to $40000$ in the dataset.
Figure~\ref{facebook-data} shows the clustering coefficient $C$, modularity
$Q$, and the number of communities of Facebook networks as a function of the
number of nodes.  As a comparison, Fig.~\ref{model-data} shows the values of
$C$ and and $Q$ from our soft-cutoff model as a function of $N$.  For the
best match between the Facebook networks and the model, it is necessary to
choose $k^*$ to be weakly increasing with $N$ and $F$ weakly decreasing with
$N$, as given in Fig.~\ref{model-data}(c).  This comparison shows that our
soft-cutoff acquaintance model can match the clustering properties of real
social networks with reasonable and slowly varying parameter values.



\section{Conclusion}

We introduced a class of acquaintance models in which macroscopic clustering
emerges naturally from the underlying social dynamics instead of
heterogeneity being one of the building blocks of the model.  Our approach is
based on distinguishing (i) direct linking, where agents with few friends
initiate connections with other agents, and (ii) transitive linking, in which
two agents become friends as a result of being introduced by a common friend.
We believe this distinction is crucial in causing the structures we see in
real world social networks.  By controlling the relative rates of direct and
transitive linking, we can generate networks that range from nearly complete,
with a tiny component of isolated cliques, to highly clustered, that are
comprised of well-defined and well-connected communities.


From the perspective of simplicity and tractability, we primarily
investigated a hard-cutoff version of the model in which only agents with
zero friends or one friend can connect to others by directly linking.  The
rate at which transitive linking occurs can be either threshold controlled or
explicitly rate controlled.  Both versions lead to highly-clustered networks
near a critical value of the relative rates of direct and transitive linking.
In the parameter regime where the long-time network is highly clustered,
there always remains a small population of marginal individuals that are part
of isolated groups. 

We also formulated and investigated a more realistic ``soft cutoff'' model,
in which the rate that an agent makes direct connections is an exponentially
decreasing function of its current number of friends.  By tuning the softness
of this cutoff, we can generate social networks whose basic geometric
features, such as the clustering coefficient, the modularity, and the number
of communities, quantitatively match those of networks from the Facebook
dataset.  Importantly, the model parameter values that are used to match with
Facebook data are fairly robust and depend quite weakly on $N$.

\bigskip\noindent We thank Mason Porter for providing the Facebook data used
in this study.  This research was partially supported by the AFOSR and DARPA
under grant \#FA9550-12-1-0391 and by NSF grant No.\ DMR-1205797.

\end{document}